\documentclass[%
 preprint,
 superscriptaddress,
 amsmath,amssymb,
 aps,
 showkeys,
 titlepage,
 endfloats*.   
]{revtex4-2}

\usepackage[utf8]{inputenc}
\usepackage[english]{babel}

\usepackage{graphicx}
\usepackage{dcolumn}
\usepackage{bm}
\usepackage{booktabs}
\usepackage[colorlinks = true,
            linkcolor = blue,
            urlcolor  = blue,
            citecolor = red,
            anchorcolor = blue]{hyperref}
\usepackage[dvipsnames]{xcolor}

\begin{document}

\title[]{First Principles study of Photocatalytic Water Splitting in BO Monolayer: 
Effect of Strain and Surface Functionalization}

\author{Soumendra Kumar Das}
\affiliation {Department of Physics, Indian Institute of Technology (Indian School of Mines) Dhanbad, 826004, India}
\author{Smruti Ranjan Parida}
\affiliation {Department of Physics, Indian Institute of Technology (Indian School of Mines) Dhanbad, 826004, India}
\author{Prasanjit Samal}
\affiliation {School of Physical Sciences, National Institute of Science Education and Research (NISER) Bhubaneswar, HBNI, Jatni-752050, Odisha, India}
\author{Brahmananda Chakraborty}
\email{brahma@barc.gov.in}
\affiliation {High Pressure and Synchrotron Radiation Physics Division, Bhabha Atomic Research Centre (BARC), Trombay, Mumbai 400085, India}
\affiliation {Homi Bhabha National Institute, Anushaktinagar, Mumbai 400094, India}
\author{Sridhar Sahu}
\email{sridharsahu@iitism.ac.in}
\affiliation {Department of Physics, Indian Institute of Technology (Indian School of Mines) Dhanbad, 826004, India}
%

\date{\today}

\begin{abstract}
 Light element based two dimensional (2D) materials are promising photocatalysts for hydrogen production via water splitting. Boron oxide (BO) is a recently synthesized 2D monolayer which has yet to be thoroughly explored for its potential applications. In this article, using first principles calculations, we report, for the first time, the visible-light photocatalytic activity of a BO monolayer for water splitting under mechanical strain and surface modification with single- and double-atom decorations (C, N, Si, Ge, P, As). The pristine BO monolayer exhibits an indirect band gap of 3.8 eV with band edges spanning the water redox potentials, but its optical absorption lies in the UV region ($\sim$4.5 eV). Strain engineering tunes the band gap and band alignment with a minimal shifting in the optical absorption ($\sim$\,0.5~\text{eV}). Single atom decoration produces a metallic state for elements like N, P, As, and an insulating state for single C, Si, Ge with a partial shifting in optical absorption. In contrast, double atom decoration produces substantial band gap reduction, improved band alignment, a pronounced red-shift in optical absorption into the visible range (1.6–3.2 eV) thus satisfying the criteria for water splitting. The stability of all the adsorbed configurations was confirmed by negative formation energy and ab-initio molecular dynamics simulations. These findings suggest BO monolayer functionalization can improve photocatalytic efficiency, providing hydrogen generation insights.
\end{abstract}

\maketitle

\section{\label{sec:level1}INTRODUCTION}
Photocatalytic water splitting is a promising technology for green hydrogen production employing solar energy, addressing the crucial energy crisis and environmental challenges forced on by the dependency on fossil fuels \cite{nishioka2023photocatalytic, chen2010semiconductor, zhang2018computational}. A photocatalyst used in water splitting must meet two requirements: one is that the sample should be semiconduting in nature with moderate energy gap, and the second is that the position of the conduction band and valence band edges must span the oxidation and reduction potentials of water \cite{li2017review, jouypazadeh2021dft}. Despite notable progress since the revolutionary work of Fujishima \textit{et al.}, the widespread application of photocatalytic water splitting remains limited by low quantum yields, inefficient visible light utilization, and insufficient catalytic activity in existing materials \cite{fujishima1972electrochemical}. Numerous investigations into water splitting on different materials, using heterogeneous catalysis, have been carried out since this groundbreaking breakthrough \cite{li2021heterojunction,  kim2018synergistic}. However, the existing photocatalytic devices are insufficiently effective and are too costly for commercial hydrogen generation. The advancement of high-performance computing has led to the design and prediction of numerous two-dimensional (2D) materials as possible photocatalysts for the water splitting process \cite{mahajan20252d}. 2D nanomaterials offer unique advantages for overall water splitting, notably high surface-to-volume ratios, short charge-carrier paths, and tunable electronic properties, all of which can enhance photocatalytic efficiency \cite{wang2020porous}\cite{singh2015computational}\cite{wu2016towards}. 

Graphene, a carbon-based 2D material, has interesting mechanical and electronic characteristics, but its zero-band gap prevents its application in photocatalysis \cite{zhang2010tio2}\cite{yang2021first}. However, other 2D materials, synthesized recently, show excllent photocatalytic properties. For example, Kishore \textit{et al.} studied nitrogenated holey graphene (C$_2$N), which has high chemical stability and an intrinsic direct band gap suitably aligned with the redox potentials for water splitting. Discovered via simple wet-chemical synthesis, C$_2$N monolayers have demonstrated promising experimental and theoretical properties for applications ranging from optoelectronics to catalysis and gas separation \cite{ashwin2017tailoring}. Das \textit{et al.} investigated the reactivity of O$_2$ and O$_3$ on the C$_2$N monolayer and reported that the 2D system exhibited resistance to ozonation with energy barriers of 0.56 eV. Interestingly, the oxidized monolayers exhibited promising potentials for water splitting applications with enhanced band edge positions and optical properties \cite{das2024unveiling}. Graphitic carbon nitride (g-C$_3$N$_4$) is an n-type semiconductor with a suitable band gap. It is an effective photocatalyst for the formation of hydrogen evolution when exposed to visible light \cite{sun2017phase}. Through first principle calculation and experimental validation, Kouser \textit{et al.} reported that 2D GaS can be an efficient photoctalyst for water splitting up to a critical thickness of 5.5 nm where the band edges straddle the water redox potentials of water.\cite{kouser20152d}. The photocatalytic activity of black phosphorus revel a tunable band gap, fast carrier mobility with optical absorption spanning from visible to infrared region \cite{yan2018recent}. The 3D/2D Graphene/MXene composite reveal exceptional photocatalytic water splitting under ultraviolet to visible region with an average hydrogen production rate of $1.4 \ \text{mmol} \ \text{h}^{-1} \ \text{g}_{\text{cat}}^{-1}$ \cite{wang2025enhanced}. Using first principle calculations and machine learning predictions, Zhang \textit{et al.} reported that the GeC/SPdSe heterostructure exhibited great performance for Z-scheme photocatalytic water splitting under visible to ultraviolet regions, achieving a maximum STH efficiency of 29.5\% \cite{zhang2025machine}. The surface modification with radicals like F and CH$_3$ significantly improved the water splitting activity of $\mathrm{CH_3@C_{60}/ZrS_2}$, $\mathrm{F@qHP\mbox{-}C_{60}/GeC}$, and $\mathrm{F@qHP\mbox{-}C_{60}/Bi}$ heterostructures with reduced Gibb's free energy and enhanced cnarge separation \cite{wanefficient}. First Principles study predicted the 2D Janus T-TiNBr structure as a stable promising photocatalyst for water splitting strong near-ultraviolet sunlight absorption and band edges staraddling the water redox potentials for HER and OER. The calculations established that visible light induced water splitting was possible at the N sites but at the Br site after surface functinalization like creating vacancy, applying external potential and adjusting the pH conditions \cite{wang2025two}. The (001) surface of ferroelectric PbTiO$_3$ exhibit promising photocatalytic activity where the water oxidation reaction follows lattice oxygen oxidation mechanism (LOM) on pristine surface but shift to the adsorbate evolution mechanism (AEM) on the surface with oxugen vacancies. The surface oxugen plays an important role in tailoring the reaction path and enabling the transition from H$_2$O$_2$ to O$_2$ production \cite{ali2025water}. After doping with C = C bonds, on the surface of hexagonal boron nitride (hBN), the hBNC/WSSe heterostructure exhiited a transition to type-I to type-II band alignment, increase in the oxidation/reduction potential and high solar hydrogen production efficiency (33.31\%) \cite{qian2023electronic}. The 2D $\beta_{2}$-$MA_{2}Z_{4}$ $(M = \text{Zr, Hf; } A = \text{Si, Ge; } Z = \text{N})$ exhibited exceptional catalytic activity for HER, OER due to synergistic effect of sliding ferroelectricity and piezoelectricity on water splitting due to the formation of type-II band alignment and enhanced interlayer vertical polarization \cite{wang2025synergistic}. The van der Waals heterostructure formed by stacking monolayers of $\mathrm{MoS_{2}}$, $\mathrm{WS_{2}}$, $\mathrm{MoSe_{2}}$, and $\mathrm{WSe_{2}}$ exhibit superior optoelectronic properties with improved photocatalytic activity due to formation of type-II band alignment, high carrier mobility, robust optical absorption in visible region and presence of strong electron-correlation effects \cite{dange2025two}. Liu \textit{et al.} reported that fabricating g-C$_3$N$_4$ with Ni$_2$P, enhances the hydrogen evolution process \cite{liu2018facile}. The C$_2$N/MoS$_2$ heterostructure exhibited improved photocatalytic water splitting under the influence of uniaxial and biaxial compressive strain \cite{das2024strain}. Priyanka \textit{et al.} studied the water splitting mechanism of RuO$_2$ and phosphomolybdic acid (PMA) caged on multi-walled carbon nanotubes. The system undergoes a metallic nature near the Fermi level. Due to the encapsulation of RuO$_2$ and PMA, the host provides effective water splitting activity \cite{bavdane2025alkaline}. RuO$_2$/IrO$_2$ are used nowadays to speed up the water splitting processes, but their low concentration and high cost make practical use extremely difficult \cite{sultan2019single}.

Some researchers have also studied the photocatalytic performance of the borate system. For instance, Wang \textit{et al.} investigated the water splitting nature of crystalline Ga$_4$B$_2$O$_9$ without any cocatalyst and reported that Ga$_4$B$_2$O$_9$ has a direct band gap of 2.16 eV and slow mobility of h+ due to the flat valence band \cite{wang2015ga4b2o9}. Similarly, Yuan \textit{et al.} reported that indium borate has the ability to photocatalyze 4-chlorophenol via photodegradation \cite{yuan2012synthesis}. Hexagonal boron nitride (hBN), a 2D material, is made up of alternating B and N atoms that are covalently bound. It has sp$^2$-hybridization like graphite, making it chemically stable and a catalyst support material \cite{gubo2018tailoring}. Meera \textit{et al.} coated hBN with Ni$_2$P and studied their hydrogen evolution nature. The modified hBN contains excess boron composites, which help in the water splitting process, with maximum hydrogen evolution activity \cite{meera2022effect}.

Interestingly, two-dimensional materials with layered architectures show great promise for dual-atom confinement. BO monolayer, a typical 2D material of lightweight, exhibits the inherent ease of controlling bimetal atoms to attain optimal catalytic activity. Mortazavi \textit{et al.} theoretically studied BO monolayers and found that it has a large indirect band gap of 3.78 eV and also low lattice thermal conductivity \cite{mortazavi2023anomalous}. However, to the best of our knowledge, no previous work has been reported on the photocatalytic activity of the BO monolayer. With this spirit, in the present work, we have analyzed, through first principles density functional theory (DFT) calculations, the structural and electronic properties of the pristine BO monolayer, along with the impact of strain and various adsorbates, aiming to enhance the photocatalytic performance of the material for water splitting.  The band structure, density of state (DOS), electron localized function (ELF), and band alignment were thoroughly examined to check the efficacy of the sample for hydrogen evolution reaction and oxygen evolution reaction.

\section{\label{sec:level2}Computational Methods}
First Principles electronic structure calculations were performed using the plane wave-pseudo potential code Vienna Ab initio Simulation Package (VASP)\cite{kresse1996efficiency}. The exchange correlation functional was parametrized by using the Perdew-Burke-Ernzerhof (PBE)~\cite{perdew1996generalized} based generalized gradient approximation (GGA). We have used the projector augmented wave (PAW) pseudopotential supplemented with Grimme’s DFT-D3 van der Waals (vdW) correction \cite{grimme2006semiempirical} to account for the long-range interaction. The cut-off value for the kinetic energy was set at 520 eV. To address the well-known underestimation of semiconductor band gaps by the PBE functional, the Heyd–Scuseria–Ernzerhof hybrid functional (HSE06) was employed for electronic structure calculations \cite{heyd2004efficient}. The crystal structure generation for the pristine and functionalized monolayer was performed using VESTA \cite{Momma2011}. A vacuum layer of 20 Å was selected along the z-direction to avoid the interaction due to the periodic images. The Brillouin zone integration was executed using a $\Gamma$-centred Monkhorst-Pack ($7 \times 7 \times 1$) \textbf{k}-mesh~\cite{monkhorst1976special} for the structural relaxation and ($10 \times 10 \times 1$) \textbf{k}-mesh for the electronic structure calculation of the pristine and adsorbed BO monolayer. The orbital projected density of states calculations were performed using tetrahedron method and a denser ($15 \times 15 \times 1$) K-mesh. For the computationally expensive HSE-06 functional, a $5 \times 5 \times 1$ K-mesh was used for the band structure calculation. All the structures were considered for the geometrical relaxations to get the optimized atomic positions unitl the total energy and the Helman-Feynman forces acting on each atom were less than 10$^{-6}$ eV and 0.01 eV/\AA~ respectively. Ab-initio molecular dynamics (AIMD) simulations have been performed with a time step of 0.5 fs for a total time period of 10 ps in the NVT ensemble where an isothermal condition is maintained using a Nosé–Hoover thermostat. The VASPKIT package was used for post-processing the computational results \cite{wang2013vaspkit}.

The position of valence band maximum and conduction band minimum was carried out with respect to the vacuum level by calculating the band gap center energy  E$_{\text{BGC}}$ and energy gap using HSE-06 functional E$_{g}^{\text{HSE06}}$ using the expression

\begin{equation}
E_{\text{CBM/VBM}} = E_{\text{BGC}} \pm \frac{1}{2} E_{g}^{\text{HSE06}}
\end{equation}

The band gap center does not get affected much by the choice of exchange correlation functional. Therefore, we have used PBE functional to calculate  E$_{\text{BGC}}$. The band edges were calculated for the pristine and adsorbed monolayer by setting their vacuum levels to zero.

\section{\label{sec:level3}RESULTS AND DISCUSSION:}

\begin{figure}[tbp!]
\includegraphics[width=1.0\linewidth]{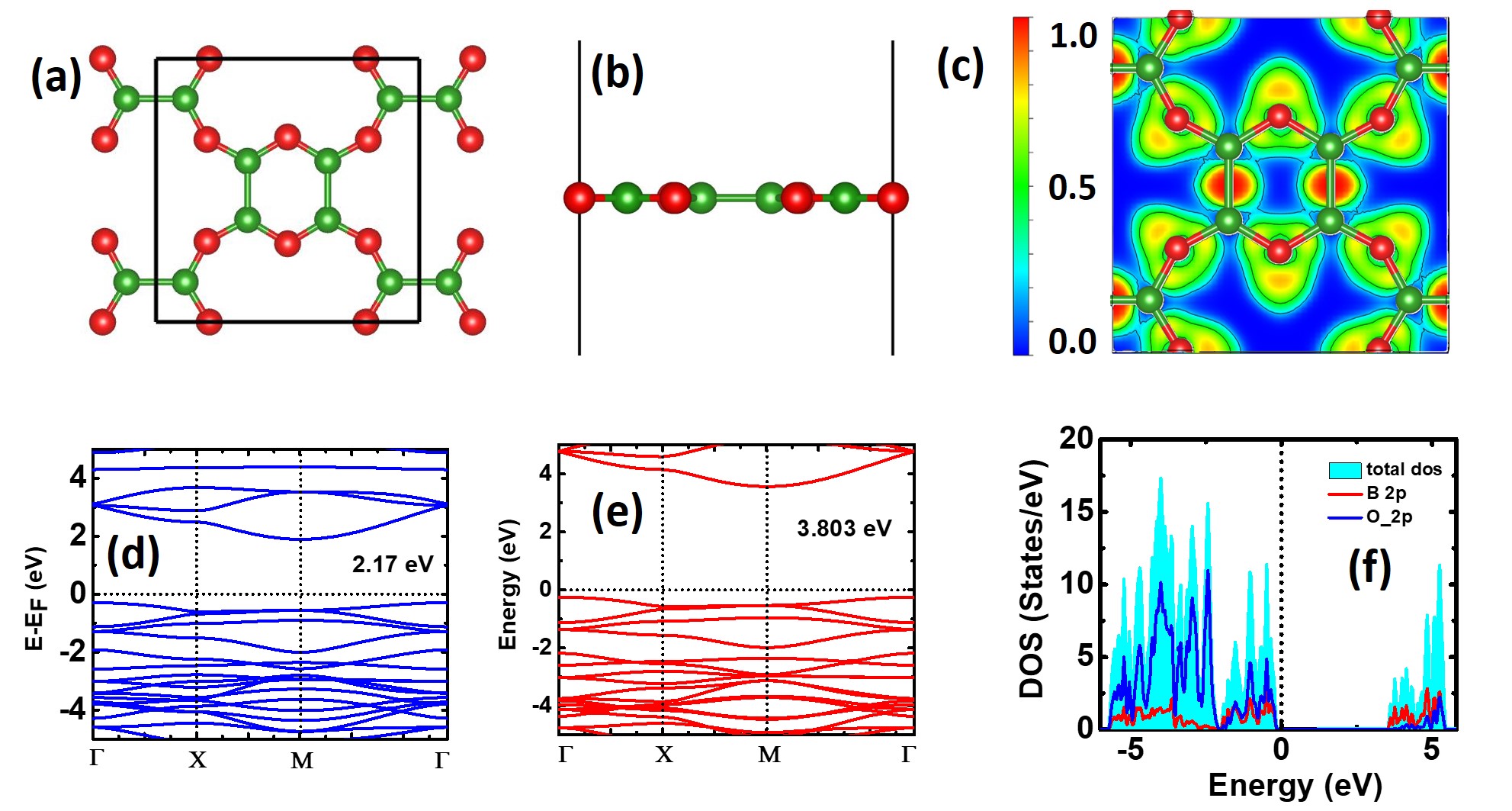}
\caption{\textbf{(a) Schematic of the crystal structure (top view), (b) side view, (c) Electronic localization function (ELF), Electronic band structure using (d) PBE-GGA, and  (e) HSE-06 hybrid functional, (f) projected density of states of pristine BO monolayer. The red and green spheres in (a,b) indicate oxygen and boron atoms, respectively.}}
         \label{pristine_BO}
\end{figure} 

The schematic representation of a single unit cell pristine BO monolayer is given in Figures~\ref{pristine_BO}(a,b). The crystal stabilizes in a square lattice with space group P4/mmm (123). The lattice constant of the optimized structure of BO is around 7.826 Å, which is close to the value 7.83 Å reported by Mortazavi \textit{\textit{et al.}}\cite{mortazavi2023anomalous}. A single unit cell of the BO monolayer contains eight boron and eight oxygen atoms connected to each other through covalent bonds. The bond length between two boron atoms (B-B) is around 1.37 Å. The boron-oxygen bond length (B-O) varies between 1.37 Å (connecting two B$_4$O$_2$ rings) and 1.39 Å (forming the B$_4$O$_2$ ring). The angle between two boron and oxygen is around $117.92^\circ$ (forming the ring) and  $146.36^\circ$ (outside the ring). The analysis of ELF gives information about the nature of the chemical bond and the distribution of electron density between two atoms. The ELF is analogous to the jellium like homogeneous electron gas and is always expressed on a scale between 0 and 1. The regions where ELF = 0, 0.5 and 1 correspond to extremely low charge density region,  fully delocalized electrons and completely localized electron region, respectively. Figure~\ref{pristine_BO}c represent the ELF plot for the pristine BO system, where the scale bar with blue, green and red colors corresponds to ELF = 0, 0.5 and 1, respectively. From the figure, it is clearly evident that the electron density is very low at boron and moderate at oxygen atoms. However, the presence of red colour indicates that the electrons are fully localized between the two boron atoms, indicating their covalent bonding nature. The calculated electronic structure of pristine BO using PBE functional (Figure~\ref{pristine_BO}d) indicates that the material is an indirect band gap semiconductor where the valence band maximum (VBM) lies at $\Gamma$ point and conduction band minimum (CBM) lies at M point. The calculated energy gap is around 2.17 eV, which is consistent with other reports \cite{othman2024light, mortazavi2023anomalous}. It is well known that the PBE functional severely underestimates the band gap and band edge positions due to the presence of derivative discontinuity and self-interaction errors. Therefore, to accurately predict the band gap, we have calculated the band structure using HSE-06 functional (Figure~\ref{pristine_BO}e), which exhibits a larger band gap around 3.8 eV consistent with previous DFT reports \cite{othman2024light, mortazavi2023anomalous}. The HSE-06 calculations preserve the nature of band dispersion obtained using the PBE functional. The conduction band minimum is dispersive in nature, indicating the presence of light electrons. Similarly, the valence band maximum is relatively flat at the $\Gamma$ point, indicating the presence of heavy holes. The nature of electronic states forming the insulating BO is analyzed through the projected density of states calculations. From the Figure~\ref{pristine_BO}f, we found that the VBM is largely populated by 2p states of oxygen, whereas the CBM is contributed by 2p states of boron. The presence of VBM close to the Fermi level indicates the p-type semiconducting nature of the BO monolayer.

\subsection{Impact of strain}
Strain is an important parameter in effectively controlling the structural, chemical and electronic properties of 2D materials. The band gap of a suitable photocatalyst should be within 1.5 eV and 3.0 eV, in order to overcome the kinetic overpotential and to avoid the electron-hole recombination. Since the pristine BO monolayer exhibits a larger band gap of 3.8 eV, it is necessary to reduce it below 3 eV in order to use the maximum fraction of visible light absorption, which constitutes 43\% of the solar spectrum. In this section, we have analyzed the electronic properties of the sample as a function of strain. The strain can be applied using the relation 

\begin{equation}
\varepsilon_{\%} = \frac{a - a_0}{a_0} \times 100\%
\end{equation}

 where $\varepsilon$ is the applied strain, a$_0$ is the lattice constant of the sample in equilibrium and 'a' is the corresponding lattice constant under strain. For a 2D monolayer system, the strain can be applied uniaxially (by varying one lattice constant keeping the other fixed) or biaxially (by varying both the lattice constants). Similarly, the strain can be applied in compression mode and tensile mode. In this work, we have studied the behaviour of the BO monolayer within $\pm 6\%$ range. The calculated energy gap of the pristine sample under uniaxial and biaxial strain using PBE and HSE-06 functional is given in Figure S1 (a,b) in the supporting information file. It is found that the energy gap variation follows a similar decreasing trend using both the PBE and HSE-06 functional. However, the magnitude of HSE-06 band gap is significantly larger in each case. It is interesting to note that compressive strain (both uniaxial and biaxial cases) increases the band gap, whereas tensile strain reduces the band gap with respect to the equilibrium configuration. The band gap decreases more rapidly with biaxial strain as compared to the uniaxial strain. Specifically for the tensile biaxial strain, the band gap decreases up to 3.27 eV. The corresponding change in the band structure is reflected in Figure \textcolor{blue} {S2}. With the application of 6\% compressive (tensile) uniaxial/biaxial strain, the CBM becomes more (less) dispersive and moves to the higher energy level (towards the Fermi level), whereas the position of VBM mostly remains unaffected. 

 The potential activity of a photocatalyst for water splitting is analyzed through the calculation of VBM and CBM with respect to the vacuum level. For a promising photocatalyst, the band edge positions should span the oxidation and reduction potentials of water. We have calculated the band edge positions of the pristine BO system under $\pm 6\%$ strain range (see Figure S3, S4 in the supporting information file). It is found that with the application of uniaxial and biaxial strain, both the CBM and VBM straddle the water redox potential values. Specifically, the VBM lies below the oxidation reaction potential (-5.67 eV) and the CBM lies above the reduction reaction potential (-4.44 eV), thus satisfying the condition for water splitting. It is interesting to mention that the CBM shows a systematic variation with respect to strain, whereas the VBM position almost remains constant. This is also consistent with the band dispersion observed in the electron structure calculations (Figure \textcolor{blue}{S2}). 

In addition to the appropriate band edge position and band gap, optical property analysis is equally important to decide whether the photocatalyst is able to absorb the maximum fraction of incoming solar radiation. The calculated optical absorption spectra of the BO monolayer under uniaxial and biaxial strain are shown in Figure \textcolor{blue}{S5} in the supporting information file. It is found that without strain, the optical absorption edge lies in the UV region (4.5 eV). Application of compressive uniaxial and biaxial strain shifts the absorption edge to higher energy, and that of tensile strain shifts to lower energy. However, this shifting is only around 0.5 eV up to $\pm 6\%$ strain. Therefore, the material can not be used for photocatalytic study under visible light.

\subsection{Impact of Surface Functionalization}
Since the application of strain could not produce a substantial improvement in the band gap, band edge positions and optical properties, we have adopted the technique of surface functionalization. A number of suitable elements, such as C, N, Si, Ge, P, As are selected for the decoration. There are three adsorption sites in the BO monolayer, such as one boron and two oxygen, as indicated in Figure \textcolor{blue}{S6} in the supporting information file. A systematic geometrical relaxation was conducted for each location, and the structure with the minimum energy was considered as the ground state. We would like to mention that the BO monolayer was analyzed with both single and double doping using these elements. It was found that, as compared to single doping, the double doping stabilized the structure with minimum energy. It is interesting to note that the surface adsorption using elements like N, P, As makes the system metallic with a finite density of states at the Fermi level (see Figure S7, S8 in the supporting informtion file). On the otherhand, single element like C, Si, Ge upon interacting with BO monolayer surface keeps the system insulating with lower band gap (see Figure S7 a,b,c in the supporting informtion file). The corresponding optical absorption in single C decorated BO monolayer undergoes an appreciable red shifting in energy but lies outside the visible region (1.6 to 3.2 eV). Therefore, we analyse the electronic properties of BO monolayer by decorating the surface with two atoms (C, N, Si, Ge, P, As) at energtically most stable adsorption sites. After adsorption, it was observed that all the decorated sample exhibit semiconducting nature with reduced energy gap. Therefore, we focus on the electronic and photocatalytic properties of the BO monolayer adsorbed with these elements, whose relaxed structure image is shown in Figure \textcolor{blue}{S9} in the supporting information file.

\begin{figure}[!tbp!]
    \includegraphics[width=1\linewidth]{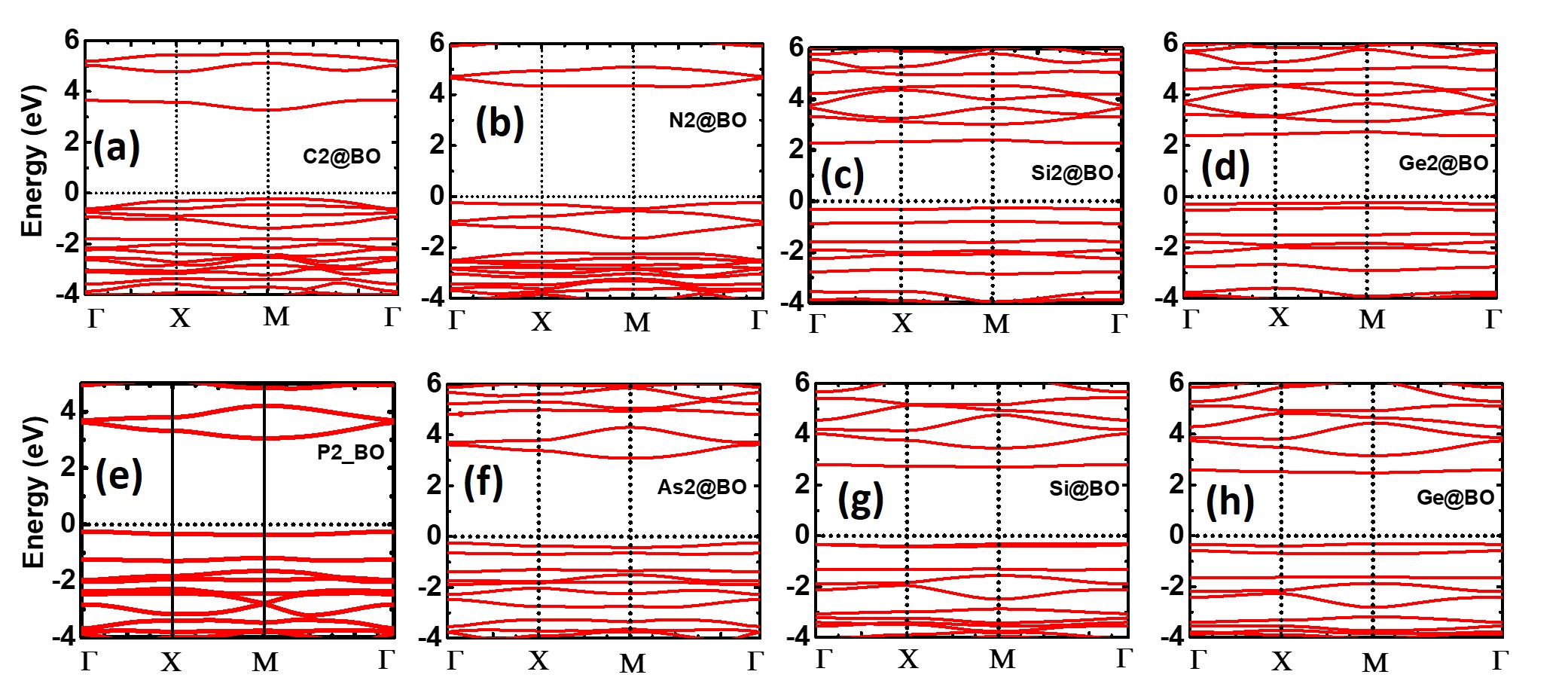}
         \caption{\textbf{(a) The calculated band structure of BO monolayer adsorbed with (a) two C, (b) two N, (c) two Si, (d) two Ge, (e) two P, (f) two As, (g) one Si, (h) one Ge atoms. The dotted horizontal line indicates the Fermi level.}}
          \label{band_hse06}
\end{figure}

Figure~\ref{band_hse06} represents the calculated band structure of the BO monolayer adsorbed with C, N, Si, Ge, and As using the HSE-06 hybrid functional. The corresponding PBE-GGA band structures are given in Figure \textcolor{blue}{S10} in the supporting information file for comparison. From Figure~\ref{band_hse06}, it was observed that after decoration with two elements of C, N, Si, Ge, P, As, and one element of Si and Ge, the band structure of all the configurations remains insulating in nature. However, the magnitude of the band gap decreases significantly compared to the pristine case.  Interestingly, for two N-decorated BO, the band gap is larger than the pristine sample. The magnitude of the band gap for all the configurations, calculated using PBE and HSE-06 functional, is given in Table 1. From  Figure~\ref{band_hse06}a, it was observed that the process of carbon decoration changes the indirect band gap structure of the pristine system to a direct band gap semiconductor. Both the CBM and VBM occur at the M point. In contrast, all the remaining adsorbates preserve the indirect band gap nature of the sample, where the CBM occurs at the M point and the VBM occurs at the $\Gamma$ point. It is further interesting to mention that for all other (C, N, Si, Ge) elements form the impurity states above the Fermi level and act as the CBM, whereas in the case of the As-decorated BO monolayer, it does not form the impurity state in the energy gap region. Although the magnitude of the band gap increases using hybrid functional, the nature of band dispersion observed in PBE and HSE-06 band structures remains almost the same.

The cohesive energy per atom for the pristine and adsorbed system is calculated through

\begin{equation}
E_{coh} = \frac{N_B E_B + N_O E_O + N_M E_M - E_{tot}}{N_B + N_O + N_M}
\end{equation}

where E$_{tot}$ is the total energy of the system, N$_{B}$, N$_{O}$, N$_{M}$ represents the number of B, O and adsorbed atoms per system and E$_{B}$, E$_{O}$ and E$_{M}$ represents the total energy of isolated B, O and adsorbed atom respectively. The calculated value of cohesive energy per atom in the BO monolayer is given in Table 1. From the table, we have found that the cohesive energy of the (Si, Ge, As) decorated monolayer decreases slightly as compared to the pristine monolayer, whereas that of C and N adsorbed BO monolayer exhibit a little higher value of cohesive energy. The increase or decrease in cohesive energy indicates the stability of the system. The slight decrease in cohesive energy indicates that the bonds are getting weaker in strength, resulting in a decrease in band gap. However, the changes are minimal in the first decimal, and the cohesive energy remains higher in values, indicating its stability.

\begin{table}[h]
\centering
\renewcommand{\arraystretch}{1.2}
\setlength{\tabcolsep}{8pt}
\begin{tabular}{lccc}
\hline
\textbf{system} & \textbf{cohesive energy per atom} (eV) & \multicolumn{2}{c}{\textbf{band gap (eV)}} \\
\cline{3-4}
& & PBE & HSE \\
\hline
BO & 7.55 & 2.17 & 3.81 \\
C2@BO & 7.71 & 1.95 & 3.49 \\
N2@BO & 7.78 & 3.18 & 4.57 \\
Si2@BO & 7.11 & 1.71 & 2.52 \\
Ge2@BO & 7.04 & 1.82 & 2.61 \\
P2@BO & 7.34 & 2.05 & 3.30 \\
As2@BO & 7.21 & 2.07 & 3.33 \\
Si@BO & 7.29 & 1.79 & 2.93 \\
Ge@BO & 7.25 & 1.63 & 2.73 \\
\hline
\end{tabular}
\caption{Cohesive energy per atom and band gap (PBE and HSE-06) for pristine and adsorbed BO monolayer.}
\label{tab:cohesive_bandgap}
\end{table}

\begin{figure*}[t!]
\includegraphics[width=1\linewidth]{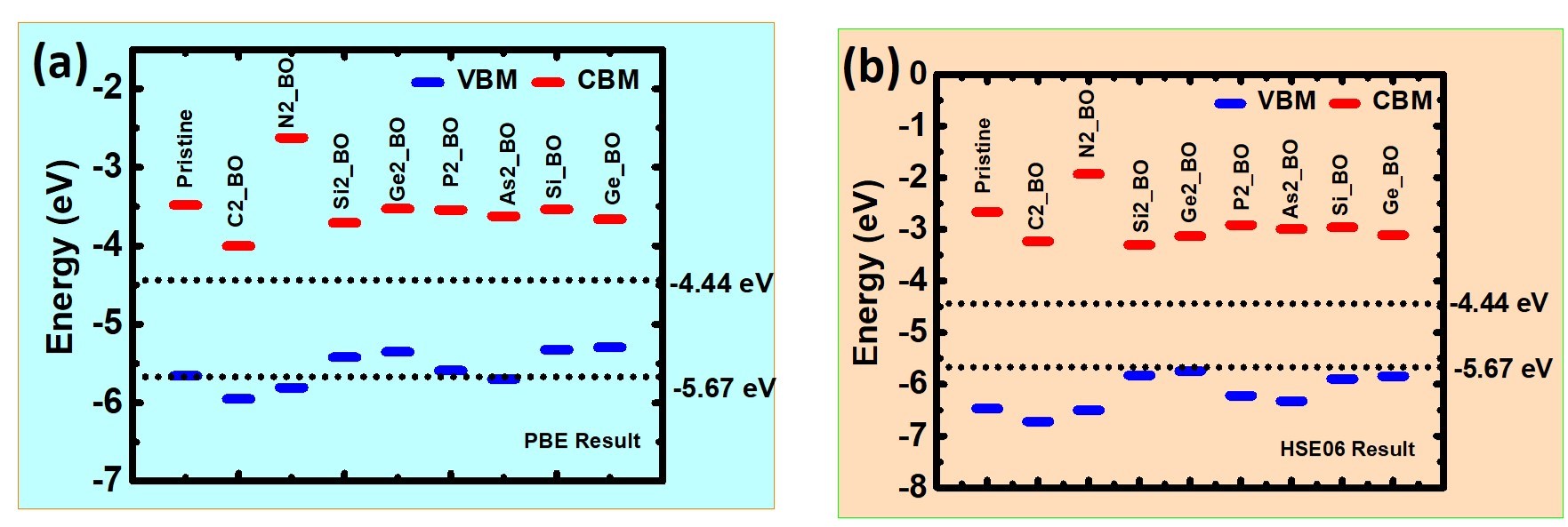}
\caption{\textbf{The band edge positions of pristine and adsorbed BO monolayer with respect to vacuum potential calculated using (a) PBE and (b) HSE-06 functional.}}
         \label{band_alignment}
\end{figure*}

The occurrence of appropriate band edge positions with respect to the hydrogen evolution reaction and oxygen reaction potential plays a crucial role in determining the suitability of a material for water splitting using solar energy. As mentioned earlier, for a promising photocatalyst, the VBM should be below the oxidation reaction potential (-5.67 eV) and the CBM should lie above the reduction reaction potential to execute the redox reactions. The calculated band edge positions for the pristine and adsorbed BO structures are displayed in  Figure~\ref{band_alignment}a. It was observed that, except for C and N, all other adsorbed structures exhibit their VBM above the oxidation reaction potential, hence not suitable for the oxygen evolution reaction. We agree that since the PBE functional could not correctly predict the band gap values, it also leads to the inaccurate prediction of band edge positions. Therefore, we have calculated the positions of the band edges using HSE-06 functional, and the results are displayed in Figure~\ref{band_alignment}b. It was found that the CBM for all the structures lie well above the reduction reaction potential, and the VBM occurs below the oxidation reaction potential. Hence, such modified structures were predicted to be thermodynamically favourable for the water splitting to occur without any external bias.

\begin{figure*}[t!]
\includegraphics[width=1\linewidth]{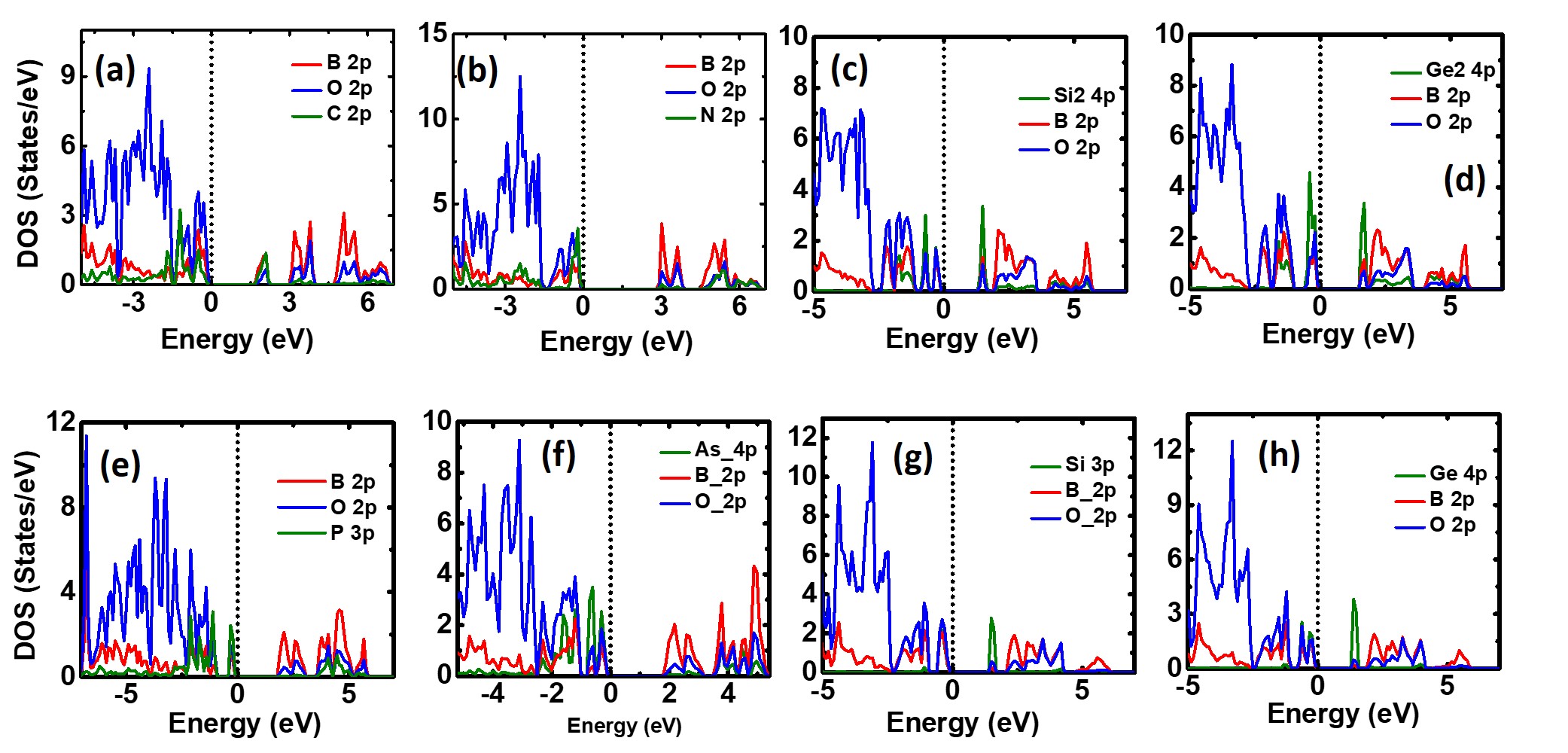}
\caption{\textbf{The calculated projected density of states of BO monolayer after decorating with (a) two C, (b) two N, (c) two Si, (d) two Ge, (e) two P, (f) two As, (g) one Si and (h) one Ge atoms. The dotted vertical line represents the Fermi level.}}
         \label{pdos_pbe}
\end{figure*}

To get further insights into the origin of band gap reduction and bonding characteristics, we have analyzed the projected density of states for the modified BO monolayer systems. From Figure~\ref{pdos_pbe}, it was observed that after decorating the surface, the CBM is mainly contributed by hybridization of p states of B, O and adsorbed atoms. For the two C decorated structure, the CBM has nearly equal contribution from C, B and O. The VBM of all the configurations has a noticeably large contribution from O 2p states. In the case of two N, two P and two As decorated BO, the contribution of the P state of N/P/As toward the CBM is negligible, and the B 2p states exhibit a high density of states. The functionalization with two Si and two Ge atoms reduces the band gap significantly, with a large population of p states of Si/Ge at the CBM. The VBM is formed by an equal population of B, O and Si/Ge 2p states. In the case of a single Si/Ge decorated BO monolayer, the CBM is mainly formed by p states of Si/Ge, whereas the VBM is dominated by both O and B atoms with a noticeable contribution from p states of O atoms. In all the cases, the VBM shows an energetically degenerate nature of p states of B, O  and ad-atoms, which indicates its covalent nature. This is also reflected in the ELF analysis of the pristine monolayer between B-B atoms.

\begin{figure*}[t!]
\includegraphics[width=1\linewidth]{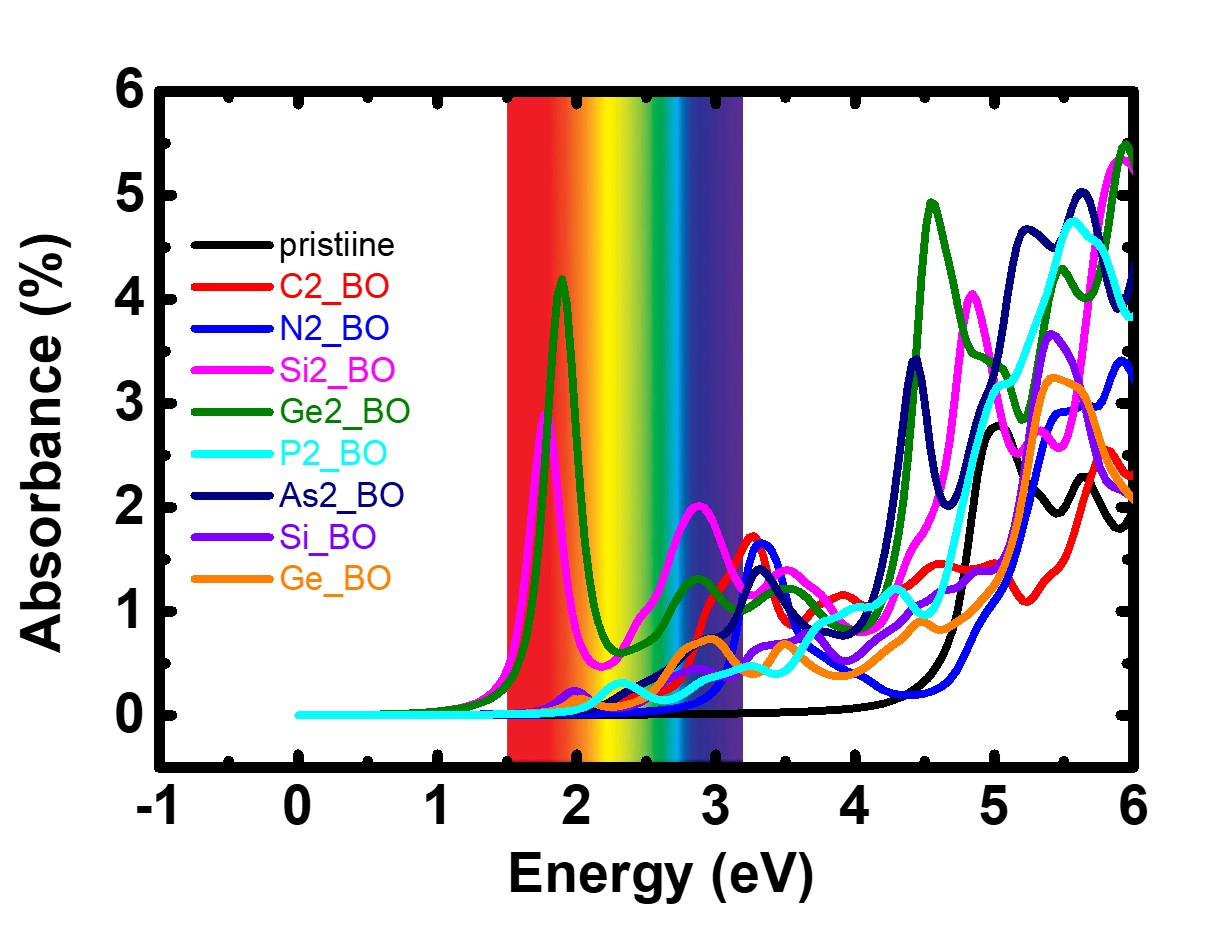}
\caption{\textbf{The optical absorption spectrum of pristine and adsorbed BO monolayer. The colored strip represents the visible region.}}
         \label{optical_pbe}
\end{figure*} 

As mentioned before, the optical property of the 2d sample also plays an important role in estimating the performance of a photocatalyst for water splitting, besides appropriate band edge positions and band gap. It determines whether the sample absorbs a large fraction of visible light to generate the electron-hole pair with efficient charge separation without recombination. The optical property of the pristine and adsorbed BO monolayer is given in Figure~\ref{optical_pbe} as a function of energy. It is found that the absorption edge, which was lying in the UV region for pristine BO, undergoes a significant red shift in energy and falls within the visible region (1.6 to 3.2 eV). Although all the decorated samples (C, N, Si, Ge, P, As) exhibited a remarkable shifting, the two Si and two Ge decorated BO exhibit the best result with the highest intensity and fall exactly around 1.6 eV, showing the best performance for water splitting. The shifting of the absorption edge to the visible region is also consistent with the reduction of the energy gap for the adsorbed monolayer.

\begin{figure*}[t!]
\includegraphics[width=1\linewidth]{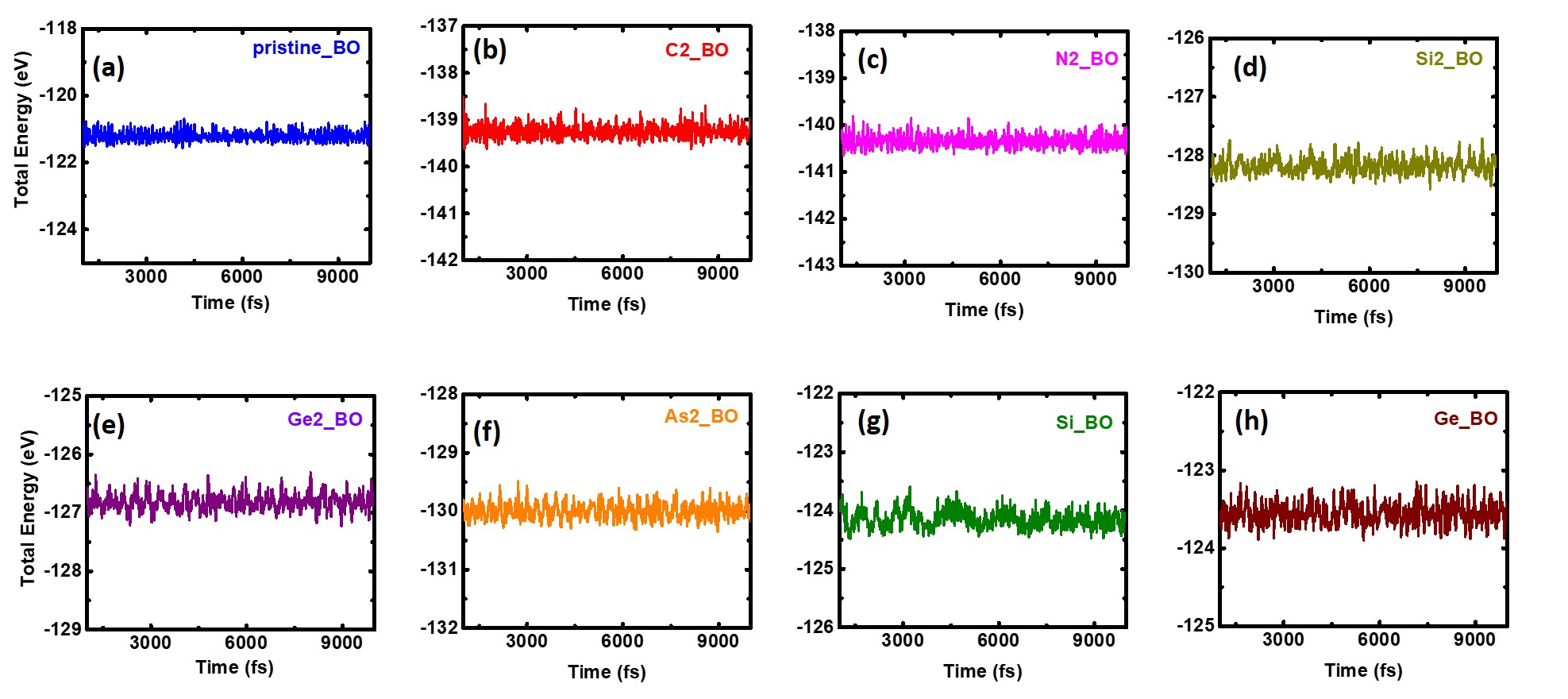}
\caption{\textbf{The AIMD simulation results showing the energy fluctuation as a function of time step for the (a) pristine and adsorbed BO monolayer with (b) two C (c) two N (d) two Si (e) two Ge (f) two As atoms and (g) single Si (h) single Ge atom}}
         \label{aimd_bo}
\end{figure*}

Despite the promising outcomes of the materials for photocatalytic applications, it is equally important to verify that the samples are stable under ambient conditions. Since DFT results are calculated at $0\,\mathrm{K}$ temperature, to check its stability at room temperature, we have carried out the AIMD simulations for the pristine and adsorbed system. The corresponding results for the fluctuation in total energy are given in the Figure~\ref{aimd_bo} and Figure S11 in the supporting information file. The plot indicates that the fluctuations in energy are minimal in both pristine and all the adsorbed structures even after 9 ps. This indicates that the samples are thermodynamically stable at room temperature making it possible for adsorption. The stability of all the samples were further analysed through the calculation of formation energy (see table S1 in the supporting information file). The expression for formation energy can be given as follows.

For single atom $X$ adsorbed on the 2D material $S$:
\begin{equation}
E_{\text{form}} = E_{S+X} - E_S - E_X
\end{equation}

For two atoms $X$ on the 2D material $S$:
\begin{equation}
E_{\text{form}} = E_{S+2X} - E_S - 2E_X
\end{equation}

For the calculation of average formation energy per adsorbate,
\begin{equation}
\bar{E}_{\text{form}} = \frac{E_{S+2X} - E_S - 2E_X}{2}
\end{equation}

where E$_{\text{form}}$, E$_{S+X}$ or E$_{S+2X}$, E$_S$, E$_X$ represents the formation energy, total energy of adsorbed system, total energy of pristine monolayer and total energy of single isolated atom respectively. Since we have analysed the sample using single and double adsorbates, we have estimated the formation energy per adsorbate for all the configurations. It was found that the formation energy of all the adsorbed system became negative which indicates the stability and feasibility of the adsorption to occur on the surface of BO monolayer.

\section{\label{sec:level4} CONCLUSION:}
In conclusion, we report the photo-catalytic activity of the BO monolayer for water splitting under the influence of mechanical strain and surface functionalization. The pristine BO exhibit an indirect band gap around 3.8 eV and band edges straddling the water redox potential values, but the optical absorption edge falls in the UV region. Under the influence of uniaxial and biaxial strain, the structure exhibits a band gap reduction in the tensile region. While the VBM position remains mostly unaltered during the entire range of strain variation, the CBM position shows variation spanning the water redox potentials. Unfortunately, the optical absorption did not show a substantial change with a minimal shifting around 0.5 eV. However, after decorating the surface with two ad-atoms like C, N, Si, P, Ge, As, etc, the monolayer exhibits a significantly reduced band gap with VBM (CBM) lying below (above) the redox potentials of water. Interestingly, the single atom (Si and Ge) decorated BO exhibits a suitable band edge for water splitting using visible light; other elements like N, P, and As exhibit metallic properties. The structural and thermal stability of the adsorbed BO monolayer were confirmed through the calculation of cohesive energy, formation energy and AIMD simulations. The optical property of the adsorbed BO indicates a substantial red shift towards the visible region. The decoration with two Si and two Ge exhibits maximum intensity, indicating its suitability as the best photocatalyst. Our calculations exhibit new insights in enhancing the photocatalytic activity of surface functionalized BO monolayer for water splitting and shed light on improving its device applications in various technological domains.

\begin{acknowledgments}
The authors would like to acknowledge the Indian Institute of Technology (Indian School of Mines), Dhanbad, for providing research facilities. The authors also acknowledge the high-performance computing facility at National Institute of Science Education and Research (NISER).
\end{acknowledgments}

\section*{Conflict of interest}
The authors have no conflicts to disclose.

\section*{Supporting Information}

The additional data that support the findings of this article are available in the Supplementary Information.

\bibliography{ref}

@PREAMBLE{
 "\providecommand{\noopsort}[1]{}" 
 # "\providecommand{\singleletter}[1]{#1}%" 
}

@article{chen2010semiconductor,
  title={Semiconductor-based photocatalytic hydrogen generation},
  author={Chen, Xiaobo and Shen, Shaohua and Guo, Liejin and Mao, Samuel S},
  journal={Chem. Rev.},
  volume={110},
  number={11},
  pages={6503--6570},
  year={2010},
  publisher={ACS Publications}
}

@article{li2017review,
  title={Review of two-dimensional materials for photocatalytic water splitting from a theoretical perspective},
  author={Li, Yunguo and Li, Yan-Ling and Sa, Baisheng and Ahuja, Rajeev},
  journal={Catal. Sci. Technol.},
  volume={7},
  number={3},
  pages={545--559},
  year={2017},
  publisher={Royal Society of Chemistry}
}

@article{jouypazadeh2021dft,
  title={A DFT study of the water-splitting photocatalytic properties of pristine, Nb-doped, and V-doped Ta3N5 monolayer nanosheets},
  author={Jouypazadeh, Hamidreza and Farrokhpour, Hossein and Momeni, Mohamad Mohsen},
  journal={Surfaces and Interfaces},
  volume={26},
  pages={101379},
  year={2021},
  publisher={Elsevier}
}

@article{fujishima1972electrochemical,
  title={Electrochemical photolysis of water at a semiconductor electrode},
  author={Fujishima, Akira and Honda, Kenichi},
  journal={nature},
  volume={238},
  number={5358},
  pages={37--38},
  year={1972},
  publisher={Nature Publishing Group UK London}
}

@article{li2021heterojunction,
  title={Heterojunction catalyst in electrocatalytic water splitting},
  author={Li, Zhenxing and Hu, Mingliang and Wang, Ping and Liu, Jiahao and Yao, Jiasai and Li, Chenyu},
  journal={Coord. Chem. Rev.},
  volume={439},
  pages={213953},
  year={2021},
  publisher={Elsevier}
}

@article{kim2018synergistic,
  title={Synergistic doping effects of a ZnO: N/BiVO 4: Mo bunched nanorod array photoanode for enhancing charge transfer and carrier density in photoelectrochemical systems},
  author={Kim, Donghyung and Zhang, Zhuo and Yong, Kijung},
  journal={Nanoscale},
  volume={10},
  number={43},
  pages={20256--20265},
  year={2018},
  publisher={Royal Society of Chemistry}
}

@article{mahajan20252d,
  title={2D nanomaterials for photocatalytic and electrocatalytic water splitting: Transforming hydrogen production for sustainable energy},
  author={Mahajan, Priyanka and Khanna, Virat},
  journal={Int. J. Hydrogen Energy},
  volume={145},
  pages={942--969},
  year={2025},
  publisher={Elsevier}
}

@article{wang2020porous,
  title={Porous two-dimensional materials for photocatalytic and electrocatalytic applications},
  author={Wang, He and Liu, Xuan and Niu, Ping and Wang, Shulan and Shi, Jian and Li, Li},
  journal={Matter},
  volume={2},
  number={6},
  pages={1377--1413},
  year={2020},
  publisher={Elsevier}
}

@article{singh2015computational,
  title={Computational screening of 2D materials for photocatalysis},
  author={Singh, Arunima K and Mathew, Kiran and Zhuang, Houlong L and Hennig, Richard G},
  journal={J. Phys. Chem. Lett.},
  volume={6},
  number={6},
  pages={1087--1098},
  year={2015},
  publisher={ACS Publications}
}

@article{wu2016towards,
  title={Towards visible-light water splitting Photocatalysts: Band engineering of two-dimensional A5B4O15 perovskites},
  author={Wu, Dihua and Zhang, Xu and Jing, Yu and Zhao, Xudong and Zhou, Zhen},
  journal={Nano Energy},
  volume={28},
  pages={390--396},
  year={2016},
  publisher={Elsevier}
}

@article{zhang2010tio2,
  title={TiO2- graphene nanocomposites for gas-phase photocatalytic degradation of volatile aromatic pollutant: is TiO2- graphene truly different from other TiO2- carbon composite materials?},
  author={Zhang, Yanhui and Tang, Zi-Rong and Fu, Xianzhi and Xu, Yi-Jun},
  journal={ACS nano},
  volume={4},
  number={12},
  pages={7303--7314},
  year={2010},
  publisher={ACS Publications}
}

@article{yang2021first,
  title={First-principles study of Graphene/ZnV2O6 (001) heterostructure photocatalyst},
  author={Yang, Anqi and Luo, Jiaolian and Xie, Zhenyu},
  journal={J. Mater. Res. Technol.},
  volume={15},
  pages={1479--1486},
  year={2021},
  publisher={Elsevier}
}

@article{ashwin2017tailoring,
  title={Tailoring the electronic band gap and band edge positions in the C2N monolayer by P and as substitution for photocatalytic water splitting},
  author={Ashwin Kishore, MR and Ravindran, Ponniah},
  journal={J. Phys. Chem. C},
  volume={121},
  number={40},
  pages={22216--22224},
  year={2017},
  publisher={ACS Publications}
}

@article{das2024unveiling,
  title={Unveiling the Reactivity of Oxygen and Ozone on C2N Monolayer},
  author={Das, Soumendra Kumar and Patra, Lokanath and Samal, Prasanjit and Sahoo, Pratap Kumar},
  journal={Phys. Status Solidi RRL},
  volume={18},
  number={12},
  pages={2400148},
  year={2024},
  publisher={Wiley Online Library}
}

@article{sun2017phase,
  title={Phase effect of Ni x P y hybridized with g-C3N4 for photocatalytic hydrogen generation},
  author={Sun, Zhichao and Zhu, Mingshan and Fujitsuka, Mamoru and Wang, Anjie and Shi, Chuan and Majima, Tetsuro},
  journal={ACS Appl. Mater. Interfaces},
  volume={9},
  number={36},
  pages={30583--30590},
  year={2017},
  publisher={ACS Publications}
}

@article{liu2018facile,
  title={Facile strategy to fabricate Ni2P/g-C3N4 heterojunction with excellent photocatalytic hydrogen evolution activity},
  author={Liu, Enzhou and Jin, Chenyang and Xu, Chenhui and Fan, Jun and Hu, Xiaoyun},
  journal={Int. J. Hydrogen Energy},
  volume={43},
  number={46},
  pages={21355--21364},
  year={2018},
  publisher={Elsevier}
}

@article{bavdane2025alkaline,
  title={Alkaline Water Splitting Over RuO2@ phosphomolybednic Acid Encased Multi-Walled Carbon Nanotubes},
  author={Bavdane, Priyanka P and Bora, Dimple K and Trivedi, Ravi Kumar and Paramasivam, Selvaraj and Ash, Anish and Nikumbe, Devendra Y and Sreenath, Sooraj and Verma, Vivek and Kuma, Shanmugam Senthil and Chakraborty, Brahmananda and others},
  journal={J. Electrochem. Soc.},
  volume={172},
  number={2},
  pages={026505},
  year={2025},
  publisher={IOP Publishing}
}

@article{sultan2019single,
  title={Single atoms and clusters based nanomaterials for hydrogen evolution, oxygen evolution reactions, and full water splitting},
  author={Sultan, Siraj and Tiwari, Jitendra N and Singh, Aditya Narayan and Zhumagali, Shynggys and Ha, Miran and Myung, Chang Woo and Thangavel, Pandiarajan and Kim, Kwang S},
  journal={Adv. Energy Mater.},
  volume={9},
  number={22},
  pages={1900624},
  year={2019},
  publisher={Wiley Online Library}
}

@article{wang2015ga4b2o9,
  title={Ga4B2O9: An efficient borate photocatalyst for overall water splitting without cocatalyst},
  author={Wang, Guangjia and Jing, Yan and Ju, Jing and Yang, Dingfeng and Yang, Jia and Gao, Wenliang and Cong, Rihong and Yang, Tao},
  journal={Inorg. Chem.},
  volume={54},
  number={6},
  pages={2945--2949},
  year={2015},
  publisher={ACS Publications}
}

@article{yuan2012synthesis,
  title={Synthesis of indium borate and its application in photodegradation of 4-chlorophenol},
  author={Yuan, Jixiang and Wu, Qiang and Zhang, Peng and Yao, Jianghong and He, Tao and Cao, Yaan},
  journal={Environ. Sci. Technol.},
  volume={46},
  number={4},
  pages={2330--2336},
  year={2012},
  publisher={ACS Publications}
}

@article{gubo2018tailoring,
  title={Tailoring the hexagonal boron nitride nanomesh on Rh (111) with gold},
  author={Gub{\'o}, Richard and V{\'a}ri, G{\'a}bor and Kiss, J{\'a}nos and Farkas, Arnold P{\'e}ter and Palot{\'a}s, Kriszti{\'a}n and {\'O}v{\'a}ri, L{\'a}szl{\'o} and Berk{\'o}, Andr{\'a}s and K{\'o}nya, Zolt{\'a}n},
  journal={Phys. Chem. Chem. Phys.},
  volume={20},
  number={22},
  pages={15473--15485},
  year={2018},
  publisher={Royal Society of Chemistry}
}

@article{meera2022effect,
  title={Effect of excess B in Ni2P-coated boron nitride on the photocatalytic hydrogen evolution from water splitting},
  author={Meera, Muraleedharan Sheela and Sasidharan, Sreekala Keerthi and Hossain, Aslam and Kiss, J{\'a}nos and K{\'o}nya, Zolt{\'a}n and Elias, Liju and Shibli, Sheik Muhammadhu Aboobakar},
  journal={ACS Appl. Energy Mater.},
  volume={5},
  number={3},
  pages={3578--3586},
  year={2022},
  publisher={ACS Publications}
}

@article{mortazavi2023anomalous,
  title={Anomalous tensile strength and thermal expansion, and low thermal conductivity in wide band gap boron monoxide monolayer},
  author={Mortazavi, Bohayra and Shojaei, Fazel and Ding, Fei and Zhuang, Xiaoying},
  journal={FlatChem},
  volume={42},
  pages={100575},
  year={2023},
  publisher={Elsevier}
}

@article{kresse1996efficiency,
  title={Efficiency of ab-initio total energy calculations for metals and semiconductors using a plane-wave basis set},
  author={Kresse, Georg and Furthm{\"u}ller, J{\"u}rgen},
  journal={Comput. Mater. Sci.},
  volume={6},
  number={1},
  pages={15--50},
  year={1996},
  publisher={Elsevier}
}

@article{perdew1996generalized,
  title={Generalized gradient approximation made simple},
  author={Perdew, John P and Burke, Kieron and Ernzerhof, Matthias},
  journal={Phys. Rev. Lett.},
  volume={77},
  number={18},
  pages={3865},
  year={1996},
  publisher={APS}
}

@article{heyd2004efficient,
  title={Efficient hybrid density functional calculations in solids: Assessment of the Heyd--Scuseria--Ernzerhof screened Coulomb hybrid functional},
  author={Heyd, Jochen and Scuseria, Gustavo E},
  journal={J. Chem. Phys.},
  volume={121},
  number={3},
  pages={1187--1192},
  year={2004},
  publisher={American Institute of Physics}
}

@article{grimme2006semiempirical,
  title={Semiempirical GGA-type density functional constructed with a long-range dispersion correction},
  author={Grimme, Stefan},
  journal={J. Comput. Chem.},
  volume={27},
  number={15},
  pages={1787--1799},
  year={2006},
  publisher={Wiley Online Library}
}

@article{monkhorst1976special,
  title={Special points for Brillouin-zone integrations},
  author={Monkhorst, Hendrik J and Pack, James D},
  journal={Phys. Rev. B},
  volume={13},
  number={12},
  pages={5188},
  year={1976},
  publisher={APS}
}

@misc{wang2013vaspkit,
  title={Vaspkit, a post-processing program for the vasp code},
  author={Wang, V},
  year={2013}
}

@article{das2024strain,
  title={Strain-Induced Enhanced Performance in 2D C2N/MoS2 Heterostructures for Photocatalytic Water Splitting: A Meta-GGA Study},
  author={Das, Soumendra Kumar and Patra, Lokanath and Samal, Prasanjit and Sahoo, Pratap K},
  journal={ACS Appl. Electron. Mater.},
  volume={6},
  number={2},
  pages={1415--1423},
  year={2024},
  publisher={ACS Publications}
}

@article{Momma2011,
  year = {2011},
  month = oct,
  publisher = {International Union of Crystallography ({IUCr})},
  volume = {44},
  number = {6},
  pages = {1272--1276},
  author = {Koichi Momma and Fujio Izumi},
  title = {VESTA 3 for three-dimensional visualization of crystal,  volumetric and morphology data},
  journal = {J. Appl. Crystallogr.}
}

@article{othman2024light,
  title={Light-metal functionalized boron monoxide monolayers as efficient hydrogen storage material: Insights from DFT simulations},
  author={Othman, Wael and Alfalasi, Wadha and Hussain, Tanveer and Tit, Nacir},
  journal={J. Energy Storage},
  volume={98},
  pages={113014},
  year={2024},
  publisher={Elsevier}
}

@article{zhang2018computational,
  title={Computational screening of 2D materials and rational design of heterojunctions for water splitting photocatalysts},
  author={Zhang, Xu and Zhang, Zihe and Wu, Dihua and Zhang, Xin and Zhao, Xudong and Zhou, Zhen},
  journal={Small Methods},
  volume={2},
  number={5},
  pages={1700359},
  year={2018},
  publisher={Wiley Online Library}
}

@article{kouser20152d,
  title={2D-GaS as a photocatalyst for water splitting to produce H2},
  author={Kouser, Summayya and Thannikoth, Anagha and Gupta, Uttam and Waghmare, Umesh V and Rao, CNR},
  journal={Small},
  volume={11},
  number={36},
  pages={4723--4730},
  year={2015},
  publisher={Wiley Online Library}
}

@article{yan2018recent,
  title={Recent progress on black phosphorus-based materials for photocatalytic water splitting},
  author={Yan, Junqing and Verma, Priyanka and Kuwahara, Yasutaka and Mori, Kohsuke and Yamashita, Hiromi},
  journal={Small Methods},
  volume={2},
  number={12},
  pages={1800212},
  year={2018},
  publisher={Wiley Online Library}
}

@article{wang2025enhanced,
  title={Enhanced Hot/Free Electron Effect for Photocatalytic Hydrogen Evolution Based on 3D/2D Graphene/MXene Composite},
  author={Wang, Qian and Ye, Zhantong and Zhao, Xiaoyu and Wang, Hejing and Zhang, Shengyan and Zhang, Suling and Liu, Hongyan and Lu, Yanhong and Jiao, Menggai and Ma, Yanfeng and others},
  journal={Small},
  pages={2407863},
  year={2025},
  publisher={Wiley Online Library}
}

@article{zhang2025machine,
  title={Machine Learning-Driven Band Alignment Strategy for Screening 1T-TMDs-Based Z-Scheme Heterostructures toward Efficient Photocatalytic Water Splitting},
  author={Zhang, Wenxue and Nie, Mengmei and He, Cheng},
  journal={Small},
  pages={2504095},
  year={2025},
  publisher={Wiley Online Library}
}

@article{wanefficient,
  title={Efficient Z-Scheme Photocatalyst for Hydrogen Production via Water Splitting Using CH3-and F-Modified C60 Fullerene-Based Heterostructures},
  author={Wan, Xue-Qing and Yang, Chuan-Lu and Shi, Wen-Jie and Li, Xiaohu and Liu, Yuliang and Zhao, Wenkai and Gao, Feng},
  journal={Small},
  pages={2504146},
  publisher={Wiley Online Library}
}

@article{nishioka2023photocatalytic,
  title={Photocatalytic water splitting},
  author={Nishioka, Shunta and Osterloh, Frank E and Wang, Xinchen and Mallouk, Thomas E and Maeda, Kazuhiko},
  journal={Nat. Rev. Methods Primers},
  volume={3},
  number={1},
  pages={42},
  year={2023},
  publisher={Nature Publishing Group UK London}
}

@article{wang2025two,
  title={Two-dimensional TiNBr as photocatalyst for overall water splitting},
  author={Wang, Yatong and Brocks, Geert and Tayran, Ceren and Er, S{\"u}leyman},
  journal={Phys. Rev. Mater.},
  volume={9},
  number={2},
  pages={025802},
  year={2025},
  publisher={APS}
}

@article{ali2025water,
  title={Water oxidation mechanisms on ferroelectric PbTi O 3 (001) surface: A density functional theory study},
  author={Ali, Ijaz and Yin, Li-Chang and Wang, Lianzhou and Liu, Gang},
  journal={Phys. Rev. B},
  volume={112},
  number={4},
  pages={045304},
  year={2025},
  publisher={APS}
}

@article{qian2023electronic,
  title={Electronic properties and photocatalytic water splitting with high solar-to-hydrogen efficiency in a hBNC/Janus WSSe heterojunction: First-principles calculations},
  author={Qian, Guo Lin and Xie, Quan and Liang, Qian and Luo, Xiang Yan and Wang, Yi Xin},
  journal={Phys. Rev. B},
  volume={107},
  number={15},
  pages={155306},
  year={2023},
  publisher={APS}
}

@article{wang2025synergistic,
  title={Synergistic effect of sliding ferroelectricity and piezoelectricity optimizes the separation of photogenerated electron and hole in two-dimensional van der Waals homostructures},
  author={Wang, Xinxin and Niu, Xianghong and Li, Gaojie and Liu, Gang and Yong, Yongliang and Li, Xiaohong},
  journal={Phys. Rev. B},
  volume={111},
  number={16},
  pages={165417},
  year={2025},
  publisher={APS}
}

@article{dange2025two,
  title={Two-dimensional transition-metal dichalcogenide--based bilayer heterojunctions for efficient solar cells and photocatalytic applications},
  author={Dange, Khushboo and Yogi, Rachana and Shukla, Alok},
  journal={Phys. Rev. Appl.},
  volume={23},
  number={1},
  pages={014008},
  year={2025},
  publisher={APS}
}

\end{document}